\documentstyle[12pt,aaspp4,graphics]{article}
\begin{document}

\title{GAMMA-RAY BURSTER COUNTERPARTS: HST BLUE AND ULTRAVIOLET DATA}
\author{Bradley E. Schaefer}
\affil{Department of Physics, Yale University, PO Box 208121, New Haven CT
06520-8121}
\affil{schaefer@grb2.physics.yale.edu}

\author{Thomas L. Cline}
\affil{Code 661, NASA/Goddard Space Flight Center, Greenbelt MD 20771}
\affil{cline@lheavx.gsfc.nasa.gov}

\author{Kevin C. Hurley}
\affil{Space Sciences Laboratory, University of California, Berkeley CA
94720}
\affil{khurley@sunspot.ssl.berkeley.edu}

\author{John G. Laros}
\affil{University of Arizona, Lunar and Planetary Laboratory, Tucson AZ
85721}
\affil{jlaros@peewee.lpl.arizona.edu}

\clearpage
\begin{abstract}

	The surest solution of the Gamma Ray Burst (GRB) mystery is to find an 
unambiguous low-energy quiescent counterpart.  However, to date no reasonable 
candidates have been identified in the x-ray, optical, infrared, or radio 
ranges.  The {\em Hubble Space Telescope (HST)} has now allowed for the
first deep 
ultraviolet searches for quiescent counterparts.  This paper reports on 
multiepoch ultraviolet searches of five GRB positions with HST.  We found no 
sources with significant ultraviolet excesses, variability, parallax, or proper
motion in any of the burst error regions.  In particular, we see no sources 
similar to that proposed as a counterpart to the GRB970228.  While this 
negative result is disappointing, it still has good utility for its strict 
limits on the no-host-galaxy problem in cosmological models of GRBs.  For most 
cosmological models (with peak luminosity $6\times10^{50}
erg \cdot s^{-1}$), the absolute B 
magnitude of any possible host galaxy must be fainter than -15.5 to -17.4.  
These smallest boxes for some of the brightest bursts provide the most 
critical test, and our limits are a severe problem for all published 
cosmological burst models.  
\end{abstract}

\section{Introduction}

	Gamma Ray Bursts (GRBs) remain the biggest mystery in modern 
astrophysics.  Much of the `blame' for this lies in the fact that no
unambiguous 
quiescent counterparts have been discovered.  Historically, source classes 
first identified outside the optical band have had to await the detection of 
counterparts before their nature was determined.  The need for the discovery 
of GRB counterparts has long been recognized, and this had led to deep searches
down to the limits of modern technology in the x-ray, optical, infrared, and 
radio bands (see Schaefer 1994 for a review).
	
	The launch and repair of the {\em Hubble Space Telescope (HST)} 
has opened up
a new window in the ultraviolet (UV) for deep imaging in GRB error boxes.  With
the exception of the extreme ultraviolet (Hurley et al. 1993), no previous work
has been published concerning searches at ultraviolet wavelengths.  There are 
many reasons to expect that a counterpart might be most visible in the UV:  (1)
Bursters might be galactic objects with an accretion disk whose hot inner edge 
will emit copious amounts of UV light. (2) Bursters might be galactic neutron 
stars with a very hot surface temperature that will primarily emit in the 
ultraviolet.  (3) Bursters might be associated with quasars or active galactic
nuclei which are characterized by UV excesses.  All these reasonable 
possibilities suggest that we should examine this new window.

	This paper reports on a search for UV counterparts to five GRBs with 
the HST.  It is the fourth in a series of papers (with similar titles) which 
report results from large observational programs aimed at detecting burst 
counterparts.  Previous papers reported deep near infrared limits for 7 GRBs 
and far infrared limits with the {\em IRAS} satellite for 23 GRBs
(Schaefer et al. 
1987), deep radio images with the {\em Very Large Array} telescope for 10
GRBs 
(Schaefer et al. 1989), and the examination of 32000 archival photographs 
for optical transients associated with 16 GRBs (Schaefer 1990).  Additional 
work has been completed on deep optical searches for counterparts to 21 GRBs 
with 300 hours of integration time.

\section{Observations}

	Gamma Ray Burst positional error regions range widely in size.  Most 
bursts (for example, those detected by the {\em BATSE} detectors on the
{\em Gamma Ray Observatory}) have typical positional uncertainties of
several degrees, while 
triangulation between widely separated spacecraft can yield up to arc-minute 
sized boxes in a few optimal cases.  Currently, only two classical GRB error 
regions have an area smaller than one square arc-minute.  These are GRB790406 
at 0.26 square arc-min and GRB790613 at 0.76 square arc-min.  These sizes
are 
to be compared to the fields-of-view of the two cameras on HST;
0.23'x0.23' for 
the FOC, and 2.5'x2.5' for the WFPC2.  Few classical GRBs can be searched
with HST if inefficient mosaics are to be avoided.

	We were granted 7.0 hours of HST time in cycles 1 and 2 (proposal 
numbers 2378 and 3984) for two GRB with the FOC camera.  The upheaval caused 
by the mirror aberration left us with the same amount of time, yet we only look
ed at GRB790113 and the field of the OT1944 (Barat et al. 1984b, Schaefer
et al. 1984, Schaefer 1990).  For HST cycle 5 (after the aberrations were 
repaired and WFPC2 was installed), we were granted 16 orbits of time (proposal 
number 5839) for the four smallest classical GRB fields.  A journal of 
observations is given in Table 1.

	A basic problem in GRB counterpart searches is that we do not know a 
{\em priori} what a burster should look like.  The many proposed burst
models 
(Nemiroff 1994) have widely disparate properties for the associated 
counterparts.  A general solution is to search for any unusual object within 
the positional error regions, the idea being that a sufficiently rare source 
is unlikely to appear inside a small box unless there is a causal connection.
Thus, counterpart searches become a statistical exercise in looking for
anomalies.  As most small boxes contain only faint sources, there are only a 
limited number of properties that can be efficiently examined.  We can look 
for unusual colors (perhaps an ultraviolet excess), variability (perhaps 
caused by changing accretion or cooling), or motion across the sky (due to 
parallax or proper motion).  A source in one of the small boxes that exhibits 
any of these properties might be sufficiently rare as to strongly argue for 
a causal connection with the burster.

	Our observational strategy was optimized to detect these anomalies.  
We looked at multiple epochs (to test for variability) separated by a half 
year (to seek proper motion) at the times of quadrature (to be sensitive to 
parallax).  We also looked in three filters so that we could construct our 
own color-color diagram from field objects.  The UV filters we chose (F195W 
and F220W for the FOC images and F218W for the WFPC2 images) were selected 
based on their throughput for UV light and the minimization of any red leaks.
These filters are broad band with FWHMs close to 40 nm and a central wavelength
as given (in nm) in the filter name.  The other two filters were the normal U 
filters (F342W for the FOC images and F336W for the WFPC2 images) and B filters
(F430W for the FOC images and F439W for the WFPC2 images).  The integration 
times were chosen to provide an equal signal-to-noise ratio for the three 
filters and an object with a Rayleigh-Jeans spectrum.  

	Full details on the detectors, filters, and calibrations are available
in various documentation provided by the Space Telescope Science Institute 
(see especially Holtzman et al. 1995).  Our data reduction has followed the 
standard pipeline processing recommended by Holtzman et al. (1995).  
In particular, we have used the PHOTFLAM data, the aperture corrections, the 
contamination corrections, and the charge-transfer-efficiency corrections.  
Our photometry likely has a systematic photometric uncertainty of a few
hundredths of a magnitude (Holtzman et al. 1995).  The median difference 
between measured magnitudes for the same star at the two epochs is 0.03 mag, 
while the standard deviation of these same measures is 0.08 mag.  Thus, 
a systematic error of $\sim 0.05$ mag should be added in quadrature to all
the statistical errors reported below.

\section{Results}

	In all cases, the sources are not seen to move from epoch-to-epoch.  
The typical positional offset between epochs is under half a WFPC2 pixel, 
where one WFPC2 pixel is 0.10''.  (An object with a transverse velocity of 
1000 $km \cdot s^{-1}$ will appear to move 0.05'' in 180 days out to a
distance of
2100 pc.)  The number of stars in our color-color diagrams is small because 
only 6 stars were visible in the UV band, and 4 of these were saturated in 
the B band.  We found no case of significant photometric variability between
epochs.  Table 2 presents the photometry for sources inside the error boxes.

\subsection{GRB790113}

	This GRB error box (Barat et al. 1984b) has a size of 78 square 
arc-minutes.  The error region for the OT1944 (Schaefer et al. 1984, 
Schaefer 1990) is 0.05 square arc-minutes in size.  This small region 
contains a faint M dwarf star which is suspected to have a time-variable
ultraviolet excess (Schaefer 1986).

	The FOC field-of-view was so small that only three stars appeared 
even in the B-band image.  One of these is the OT1944 candidate star, while 
the other two were a pair of faint stars located roughly 15'' to the east.  
With a ground based calibration of the brightness of the star pair, the 
brightness of the candidate is $B=22.8 \pm 0.1$.  Seven ground-based
brightness 
measures for this source from 1983 to 1994 show B magnitudes ranging from 
23.06 to 23.48 (with typical error of 0.11 mag) and one measure of $22.65
\pm 0.3$.  In the U-band, only the star pair was detected, while the
candidate had a 5-sigma detection limit of $\sim 24.0$ for all epochs.  In the
UV-band, no 
source was ever detected at any epoch, with a typical 5-sigma threshold of 
$\sim 22.1$ mag.  Geometric distortions and the lack of background stars
prevented the co-adding of images from different epochs.

\subsection{GRB790325}

	This GRB error box has a size of 2 square arc-minutes (Laros et al. 
1985).  The bright star 104 Herculis and the OT1946 (Hudec et al. 1987; Hudec,
 Peresty, \& Motch 1990) are both nearby, but significantly outside the
error 
box.  No deep optical studies of this field have been published.

	Over the whole field-of-view, the HST B-band images show 25 sources, 
the U-band images show 15 sources, while the UV-band images show only 1 source.
Of the 25 sources, three are galaxies.  The WFPC2 field-of-view completely
covers the GRB box.  Eight sources are inside the GRB error region, of
which one is a galaxy (see Figure 1).  The brightest source in the box is a 
G-type star that was saturated in the B-band, and had magnitude
$14.50 \pm 0.01$ in 
the U-band, and $16.77 \pm 0.01$ in the UV-band.  The only galaxy in the
box has
magnitudes $B=21.98 \pm 0.05$ and $U=23.23 \pm 0.37$.  The limiting
magnitudes for a 
5-sigma detection are 23.0 in the B-band, 22.4 in the U-band, and 20.0 in the 
UV-band.

\subsection{GRB790406}

	This burst has by far the smallest of all classical GRB error boxes 
(Laros et al. 1981).  Previous optical studies have been published by 
Chevalier et al. (1981) and Motch et al. (1985).

	Over the whole field-of-view, the HST B-band images show 10 sources, 
the U-band images show 9 sources, while the UV-band images show 2 sources 
(both near the limits of detection).  Of these 10 sources, four are galaxies.
None of these sources is inside or near the GRB error region.  The limiting 
magnitudes for a 5-sigma detection are 22.8 in the B-band, 22.5 in the U-band,
and 20.0 in the UV-band.

\subsection{GRB790613}

	This burst has the second smallest of all classical GRB boxes (Barat 
et al. 1984a).  Optical studies appear in Ricker, Vanderspek, \& Ajhar
(1986),
 Vrba, Hartmann, \& Jennings (1995, VHJ), and Sokolov et al. (1995,
SKZKB).

	Over the whole field-of-view, the HST B-band images show 13 sources, 
the U-band images show 7 sources, while the UV-band images show 2 sources 
(both near the limits of detection).  Of these 13 sources, five of them are 
galaxies.  Three sources appear inside the GRB error box, and all three of 
these are galaxies.  The first source (corresponding to object \#63 in VHJ
and
 galaxy 3 in SKZKB)has $B=22.05\pm0.05$ and $U=21.84\pm0.07$.  The second
galaxy in 
the GRB box (corresponding to object \#71 of VHJ and galaxy 2 in SKZKB)
has 
$B=22.37\pm0.07$ and $U=22.83\pm0.16$.  The third galaxy (corresponding to
object
\#57 
of VHJ and object b in SKZKB) has $B=22.33\pm0.06$ and $U=22.09\pm0.15$.
The limiting 
magnitudes for a 5-sigma detection are 23.2 in the B-band, 22.7 in the U-band, 
and 20.4 in the UV-band.

	The GRB box is long and thin, and the extreme edges are outside the 
WFPC2 field-of-view.  Close to 10\% of the error box is not covered by the
HST 
data; for this excluded portion, ground based images in the B- and R-bands do 
not show any sources that are bright enough to be expected to be visible.  
Thus, for galaxies, the limits on the brightest galaxy in the GRB790613 box 
are $B=22.05\pm0.08, U=21.84\pm0.07$, and $UV=20.4$.

\subsection{GRB920406}

	This GRB error box has a size of 2 square arc-minutes.  No optical 
studies of this field have been published.

	Over the whole field-of-view, the HST B-band and U-band images show 25
sources, while the UV-band images show only 1 source.  Of these 25 sources, 
two are galaxies.  The WFPC2 field-of-view misses the extreme tips of the GRB 
box, although 85\% coverage is attained.  Eight sources are inside the GRB
error region, of which none is a galaxy (see Figure 2).  The limiting magnitude
s for a 5-sigma detection are 23.0 in the B-band, 22.6 in the U-band, and 19.5 
in the UV-band.

\section{Discussion}

	HST opens up a new window in the ultraviolet for GRB quiescent 
counterparts searches.  We might expect to see such sources since bursters 
might contain hot accretion disks, hot neutron stars, or the blue cores of 
active galactic nuclei.  Our search of five GRB error regions includes the 
four smallest classical GRB boxes.  We have found no sources with any unusual
property, including UV excess, variability, parallax, or proper motion.  
This lack of counterparts is disappointing.  With these first UV searches, 
the entire accessible electromagnetic spectrum has now been examined for 
quiescent counterparts.

	This lack of counterparts is nevertheless critical for several 
topics of recent interest.  First, the GRB970228 has recently been associated 
with a fading x-ray transient which has been associated with an optical 
transient which has been associated with a faint point source superposed on 
an extended source (see IAU Circulars from numbers 6572 to 6631 for references)
.  While each of these associations can be questioned and the various 
implications are currently unclear, the widely publicized interpretation is 
that the point source plus extended source is the quiescent counterpart.  
On the assumption that this identification is correct, we can examine whether 
similar counterparts are visible in our HST data.  If the counterparts are 
similar to the GRB970228 candidate (B=25.4), then the expected brightness 
should scale as the GRB's peak flux.  GRB970228 has a peak flux of roughly 
$1\times10^{-6} erg\cdot cm^{-2}\cdot s^{-1}$, while the four classical
GRBs have peak
fluxes as listed in Table 4.  The calculated B magnitudes for a counterpart
like GRB970228 (see Table 3) were corrected for the galactic extinction as 
prescribed by Burstein \& Heiles (1982).  For comparison, Table 3 also
lists 
our HST limits for extended sources and for extended sources associated with 
point sources.  We see that any counterpart as proposed for GRB970228 should 
be easily detected in the boxes of GRB790406 and GRB790613.  This could 
alternatively be viewed as evidence against the identification of the 
GRB970228 candidate, evidence for a large spread in luminosity, or as evidence
that both the extended and point source counterparts fade substantially on a 
time scale shorter than years.

	Second, the presence of only very faint galaxies in these smallest of 
boxes is a strong challenge to most cosmological models of bursts.  Let us take
 the peak luminosity to be $6\times10^{50} erg\cdot s^{-1}$ (e.g.,
Fenimore et al.1993) 
which is adopted for virtually all cosmological models due to the
agreement with the LogN-LogP curves, the reported time dilation, and neutron 
star energetics.  The distances can then be calculated for each burster based 
on their observed peak fluxes (see Table 4).  The distance to the burster must
be the same as the distance to the host galaxy.  Then, the derived distance and
our HST limit on the apparent magnitude of any extended source in the GRB 
region directly yield a limit on the absolute magnitude for the host galaxy.  
This limit must be corrected for the extinction through our galaxy, for which 
we have used the prescription in Burstein \& Heiles (1982).  The blue
absorption
is less than 0.3 mag in all cases, and this is confirmed by the lack of far 
infrared cirrus in the regions.  In Table 4, we present our limits on the 
absolute magnitude for the host galaxies.

	The host galaxies must therefore be fainter than absolute B magnitudes
from -15.5 to -17.4.  This is to be compared with the values for an L* galaxy
of -21.0 and for a small irregular galaxy (like the SMC) of -16.2.  Here, we 
have four-out-of-four GRBs that can have hosts no brighter than irregular 
galaxies.  Such faint galaxies occupy the lowest $\sim 2\%$ of the galaxy
luminosity 
function.  The probability of getting such a result is $\sim 10^{-7}$ if
the GRB hosts are drawn from a normal selection of galaxies.  With 
four such limits, it is difficult to invoke a significant width to the 
luminosity function as an explanation.

	Our result from HST by itself presents a no-host-galaxy dilemma for all
cosmological models.  Thus, now any acceptable cosmological model must explain
the lack of host galaxies to deep limits.  Only two potential solutions exist:
(1) Bursts might be placed at such a great distance that the host is very faint
.  But to do this denies the observed dilation results, violates energy 
availability even for the annihilation of a whole neutron star, and forces a 
contrived evolution to explain the -1.5 slope of the bright portion of the 
LogN-LogP curve.  (2) Bursts might be required to occur outside host galaxies. 
But then it is unclear why bursters exist only in intergalactic space, and most
models require a galactic environment.  We are not aware of any published 
cosmological model which has yet successfully solved the no-host-galaxy dilemma.

Support for this work has come under grants from the Space Telescope Science 
Institute (numbers 2378 and 5839) as well as NAG 5-1560 and JPL-958056.

\clearpage
\begin{table}
\begin{tabular}{ccccc}

Date            & Detector & Target & Filters & Exposure\\        
\hline
1991 Oct 3	&FOC	& GRB790113	& UV, U	        & 1782 s\\
1992 Mar 14	&FOC	& GRB790113	& UV, U		& 1792 s\\
1992 Mar 20	&FOC	& GRB790113	& UV, U, B	& 5000 s\\
1992 Sep 8	&FOC	& GRB790113	& UV, U, B	& 2694 s\\
1995 Jul 8	&WFPC2	& GRB790613	& UV, U, B	& 5000 s\\
1995 Sep 1	&WFPC2	& GRB790325	& UV, U, B	& 4200 s\\
1995 Sep 6	&WFPC2	& GRB920406	& UV, U, B	& 4400 s\\
1995 Oct 30	&WFPC2	& GRB790406	& UV, U, B	& 4400 s\\
1995 Dec 29	&WFPC2	& GRB790613	& UV, U, B	& 5000 s\\
1996 Feb 27	&WFPC2	& GRB790325	& UV, U, B	& 4200 s\\
1996 Mar 4	&WFPC2	& GRB920406	& UV, U, B	& 4400 s\\
1996 Apr 17	&WFPC2	& GRB790406	& UV, U, B	& 4400 s\\     	
                &	&		&	&13.1 hour\\
\end{tabular}
\caption{Journal of HST GRB observations.}
\end{table}

\clearpage
\begin{table}
\begin{tabular}{cccccc}

	& Object &&&&\\
Region          &	Designation	& B      &U      &UV &Comments\\
\hline
GRB790113	& G	& 22.8	& $>$24.0	& $>$22.1	& M star,
variable in B?\\
GRB790325	& A	& (sat)	& 14.50	& 16.77	& G star\\
	& B	& 18.87	& 20.48	& $>$20.0 &\\
	& C	& 20.61	& $>$22.4	& $>$20.0 &\\
	& D	& 21.98	& 23.23	& $>$20.0	& Galaxy\\
	& E	& 18.37	& 19.23	& $>$20.0	&\\
	& F	& 20.27	& $>$22.4	& $>$20.0 &\\  
	& I	& 20.62	& $>$22.4	& $>$20.0 &\\
	& J	& 20.12 & 20.47	& $>$20.0 &\\
GRB790406	&...	& $>$22.8	& $>$22.5	& $>$20.0	&
Region is empty\\
GRB790613	& VHJ\#63 &22.05	& 21.84	& $>$20.4	& Galaxy\\
	& VHJ\#71	& 22.37	& 22.83	& $>$20.4	& Galaxy\\
	& VHJ\#57	& 22.33	& 22.09	& $>$20.4	& Galaxy\\
GRB920406	&A	& (sat)	& 15.99	& $>$19.5 &\\
	& B	& 18.95	& 19.83	& $>$19.5 &\\
	& E	& 20.30	& 20.82	& $>$19.5 &\\
	& I	& 16.41	& 17.37	& $>$19.5 &\\
	& J	& 17.36	& 19.15	& $>$19.5 &\\
	& L	& 17.45	& 18.24	& $>$19.5 &\\
	& Y	& 20.52	& 21.29	& $>$19.5 &\\
	& Z	& 19.32	& 19.69	& $>$19.5 &\\
\end{tabular}
\caption{Sources inside error regions.  }
\end{table}

\clearpage
\begin{table}
\begin{tabular}{cccc}

	& Expected B for candidate	& B$_{lim}$ for		&
B$_{lim}$ for extended\\
Region  & like for GRB970228	& extended sources & plus point sources\\      
\hline
GRB790325	& 22.0	& $\geq$ 21.98	& $>$23.0\\
GRB790406	& 21.2	& $>$ 22.8	& $>$22.8\\
GRB790613	& 20.9	& $\geq$ 22.05	& $>$23.2\\
GRB920406	& 23.2	& $>$ 23.0	& $>$23.0\\
\end{tabular}
\caption{Expected brightness of counterpart like proposed for GRB970228.}
\end{table}

\clearpage
\begin{table}
\begin{tabular}{ccccc} 

	& Peak Flux & Distance$^{a}$ & Galactic & $M_{B}^{o}$ for\\
Region  & $(erg\cdot cm^{-2}\cdot s^{-1})$	& (Mpc) & latitude & host
galaxy\\
\hline
GRB790325	& $1.5\times10^{-5}$	& 580	& $22^{o}$ &$\geq -17.2$\\
GRB790406	& $2.4\times10^{-5}$	& 460	& $-61^{o}$&$>-15.5$\\
GRB790613	& $3.2\times10^{-5}$	& 400	& $38^{o}$ &$\geq -16.0$\\
GRB920406	& $4.6\times10^{-6}$	& 1050	& $-28^{o}$&$>-17.4$\\
\end{tabular}
\caption{No-host-galaxy limits for the four classical GRB regions.
$^{a}$ This assumes a peak luminosity of $6 \times 10^{50} erg \cdot 
s^{-1}$ as adopted by virtually all cosmological
models of GRBs.}
\end{table}

\clearpage
\begin{figure}
\caption{B mosaic for GRB790325.
This WFPC2 image through the F439W filter shows no bright galaxies, as expected
 if bursters are in host galaxies.  The only galaxy visible inside the error 
box is object D, at B=21.98.  The diagonal streak is from the star 104 Her.
}
\end{figure}

\begin{figure}
\caption{
B mosaic of GRB920406.
The error box extends to just outside the WFPC2 field of view, such that the
observations cover 85\% of the box.  (Ground based images show no bright
sources
in the missing regions.)  All objects inside the box are point sources, not 
galaxies.
}
\end{figure}
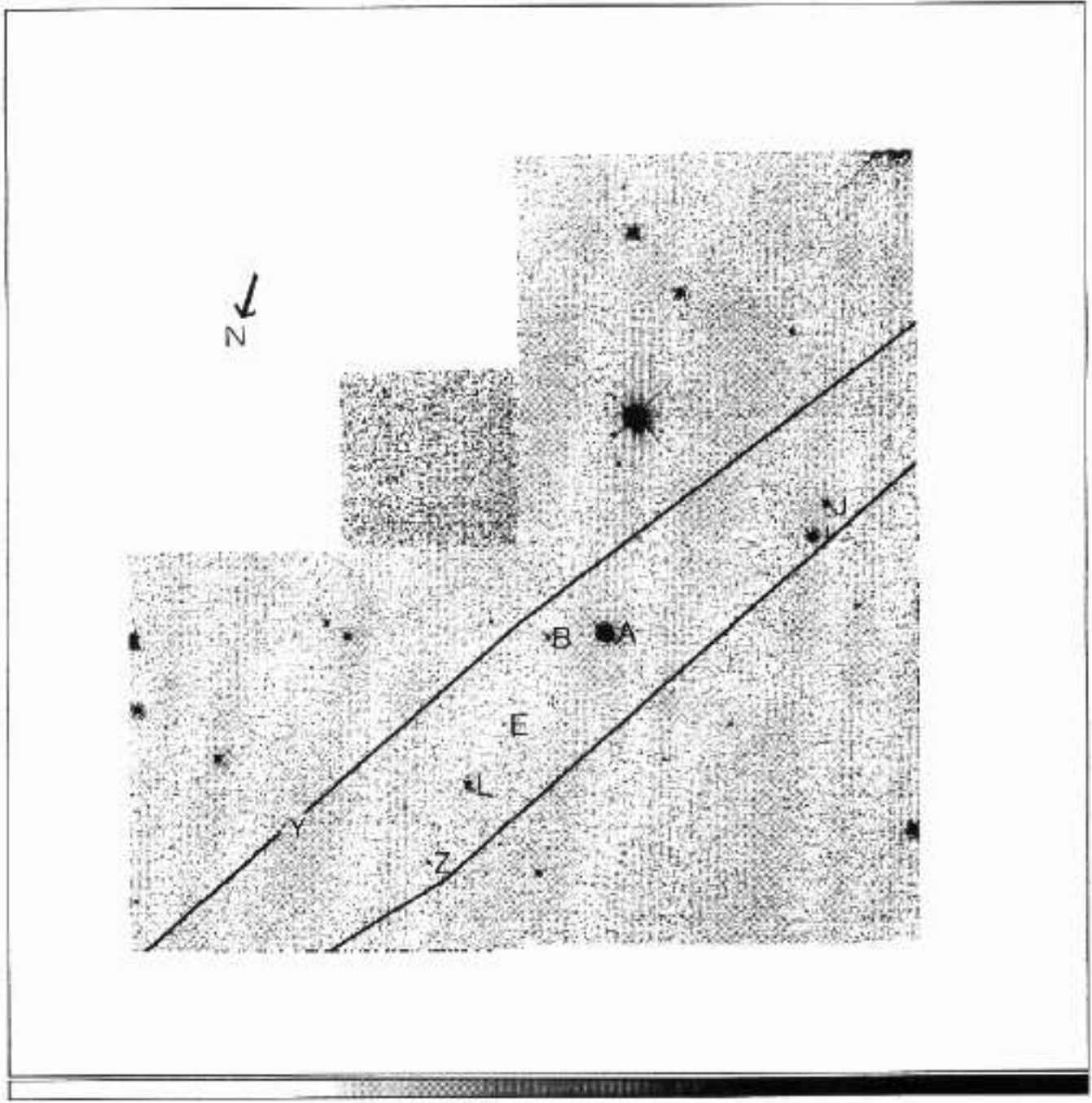
\end{document}